\documentclass[a4paper]{article}

\usepackage{INTERSPEECH2021}
\usepackage{caption}
\usepackage{subcaption}
\usepackage{multirow}
\usepackage{cite}
\usepackage{setspace}
\usepackage{enumitem}

\title{Listen with Intent: \\Improving Speech Recognition with Audio-to-Intent Front-End}
\name{
\begin{tabular}{@{}c@{}}
Swayambhu Nath Ray$^{2\star}$,
Minhua Wu$^{1\star}$,
Anirudh Raju$^{1\star}$,  
Pegah Ghahremani$^{1\star}$, 
Raghavendra Bilgi$^{2\star}$, \\
Milind Rao$^{1}$, 
Harish Arsikere$^{2}$,
Ariya Rastrow$^{1}$,
Andreas Stolcke$^{1}$,
Jasha Droppo$^{1}$
\end{tabular}}
\address{
$^{1}$Amazon Alexa, USA {\hspace*{0.5cm}} $^{2}$Amazon Alexa, India
\thanks{* Equal contribution}}
\email{\{swayar, wuminhua, pegahgh, ranirudh, rrbilgi\}@amazon.com}

\begin{document}

\maketitle
\begin{abstract}
\label{sec:abstract}
Comprehending the overall intent of an utterance helps a listener recognize the individual words spoken. Inspired by this fact, we perform a novel study of the impact of explicitly incorporating intent representations as additional information to improve a recurrent neural network-transducer (RNN-T) based automatic speech recognition (ASR) system. An audio-to-intent (A2I) model encodes the intent of the utterance in the form of embeddings or posteriors, and these are used as auxiliary inputs for RNN-T training and inference. 
Experimenting with a 50k-hour far-field English speech corpus, this study shows that when running the system in \textit{non-streaming} mode, where intent representation is extracted from the entire utterance and then used to bias streaming RNN-T search from the start, it provides a 5.56\% relative word error rate reduction (WERR). On the other hand, a \textit{streaming} system using per-frame intent posteriors as extra inputs for the RNN-T ASR system yields a 3.33\% relative WERR. A further detailed analysis of the streaming system  indicates that our proposed method brings especially good gain on media-playing related intents (e.g. 9.12\% relative WERR on PlayMusicIntent).
\end{abstract}

\section{Introduction}
\label{sec:intro}
Spoken language understanding (SLU) systems are conventionally designed as a pipeline that includes an automatic speech recognition (ASR) system that converts speech to text, followed by a natural language understanding (NLU) system that extracts structured data such as domain, intent and slots.

For the ASR system,  end-to-end models have gained popularity in recent years as they combine separate components of conventional DNN-HMM hybrid ASR systems \cite{hinton2012deep} (acoustic, pronunciation and language models) into a single neural network. End-to-end models include connectionist temporal classification \cite{graves2006connectionist}, recurrent neural network-transducer (RNN-T) \cite{graves2012sequence}, and attention-based sequence-to-sequence models \cite{bahdanau2014neural,chorowski2015attention,ray2018ad3} also known as LAS: Listen, Attend and Spell \cite{chan2016listen}. Among these three methods, RNN-T is replacing the traditional hybrid ASR models \cite{he2019streaming,sainath2019two} since it has good streaming capability which is challenging to LAS and does not have CTC's frame-independence assumption. Various directions have been explored to enhance RNN-T ASR performance. Depth-LSTM and layer normalization was tried in \cite{li2019improving}. Using LAS as a second-pass rescorer by attending to both encoder features and n-best output from the RNN-T has been explored in \cite{sainath2019two}, and an inter encoder-decoder attention mechanism was introduced in \cite{wang2020attention} to better align the encoder feature with the hypothesis. Other improvements include better initialization methods, training on TTS data, and use of lookahead encoders \cite{li2020developing}. All these improvements on the RNN-T ASR system so far have focused on better acoustic or language modeling and rescoring extensions without adding any capability of understanding to the system.

When humans process speech signals, transcription and understanding happen simultaneously, and the capability of understanding enables human to usually provide better quality of transcription than machines. Training an end-to-end SLU system to predict intent and slot values directly from audio is therefore becoming a popular research area \cite{qian2017exploring,serdyuk2018towards,chen2018spoken,haghani2018audio, lugosch2019speech,rao2020speech,rao2021mean}. While such system may not readily outperform or replace large-scale  conventional SLU systems with independently optimized ASR and NLU modules, it is indicated that semantic information such as intent and slots could potentially help improve an ASR system \cite{haghani2018audio,rao2021mean}. However, these approaches were incorporating semantic information implicitly by using training intent and slot prediction as extra tasks. The technique to explicitly incorporate various contextual signals analogous to intent, such as dialog state and music play state, into an RNN-T based ASR system has been proposed in \cite{wu2020multistate,ray2021improving}, but most of these contextual signals are derived only after the first turn of the dialog and would only benefit subsequent turns. Therefore, extra studies are needed to investigate the impact of using semantic embeddings produced on-the-fly directly with input audio features on an ASR system. Focusing on intent-based semantic embeddings, our contributions in this study are as follows:
\vspace{-1mm}
\begin{itemize}[leftmargin=5mm]
    \item We propose to incorporate intent embeddings of the audio into the RNN-T ASR system to improve its recognition accuracy through an auxiliary audio-to-intent (A2I) front-end.
    \item We run extensive experiments on a large corpus of 50k hours of far-field US English speech to demonstrate effectiveness of the proposed approach.
\end{itemize}
To the best of our knowledge, this is the first study to enhance RNN-T ASR performance using intent representations produced on-the-fly from an auxiliary intent prediction model.

\vspace{-1mm}
\section{Method}
\label{sec:approach}
\subsection{RNN-T ASR using intent representations}
\label{sec:a2i-asr-integrate}
We train a new RNN-T ASR system where the input features for the encoder contain frames of original audio feature vectors along with intent representations obtained by feeding the audio frames into a pre-trained audio-to-intent (A2I) model which will be described with more details in Section \ref{sec:A2I model}. For results described in this paper, we incorporate intent representations only into the encoder portion of the RNN-T, since our preliminary experiments showed that feeding extra embeddings to the prediction network is less effective. (This aligns with the findings in \cite{ghodsi2020rnn} that RNN-T encoder makes better use of contextual information than the prediction network).

\begin{figure}[h]
\centering
  \includegraphics[width=0.7\linewidth]{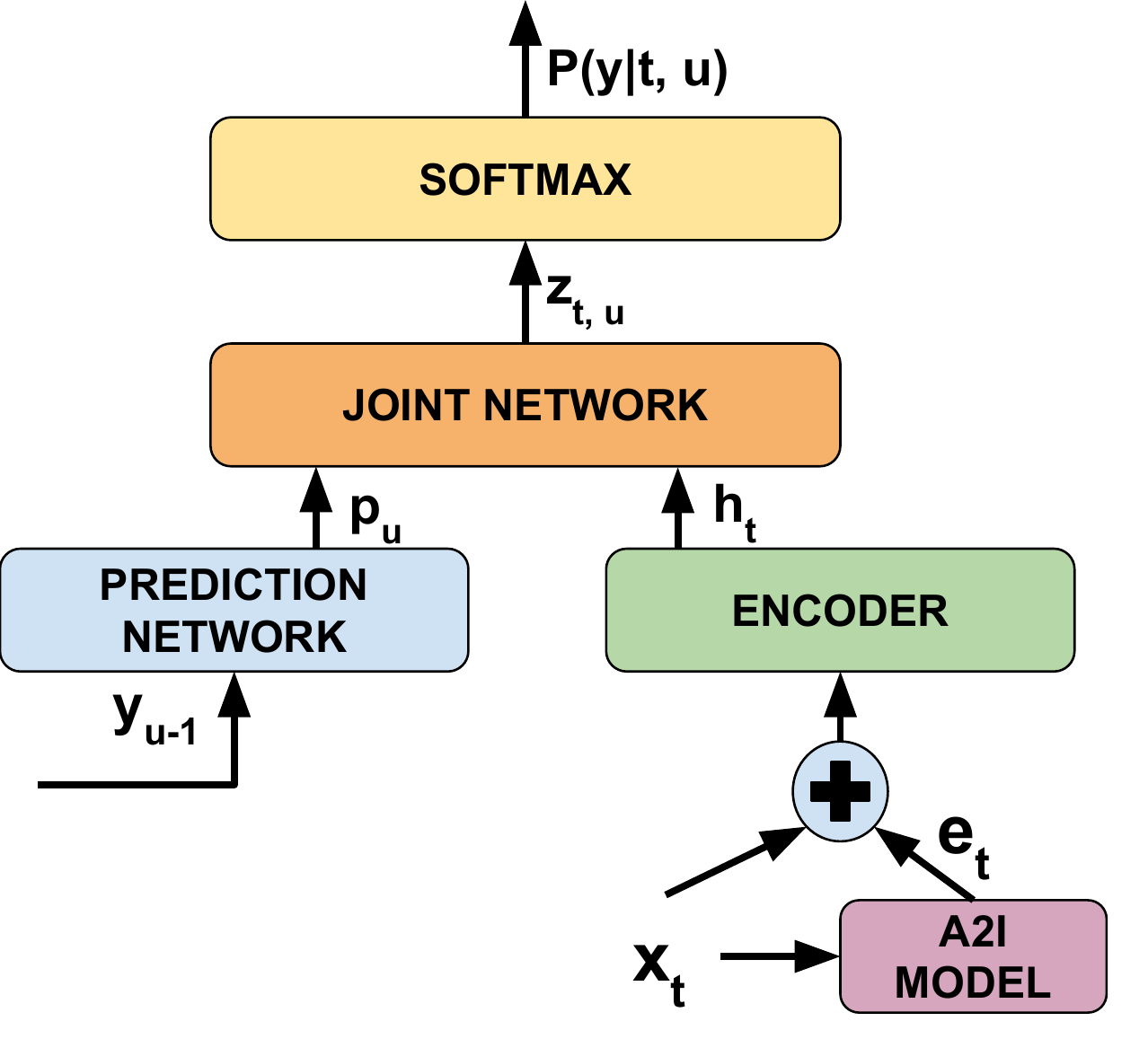}
  \caption{\small{\textit{Incorporate intent embeddings into RNN-T ASR in a streaming fashion. Audio features at each frame ($\text{x}_{t}$) are concatenated with per-frame intent embeddings ($\text{e}_{t}$) inferred from a pretrained A2I model.}}}
  \label{fig:incorporate-intent-representation}
\end{figure}

Specifically, we concatenate each frame of input audio features ($\text{x}_{t}$) with its intent embeddings ($\text{e}_{t}$) (as illustrated in Figure \ref{fig:incorporate-intent-representation}) to train the proposed RNN-T ASR model. The A2I model is pre-trained from a much smaller intent-annotated corpus, and its parameters are not updated during RNN-T ASR model training. For a streaming ASR system, per-frame intent embeddings ($\text{e}_{t}$) are required to be concatenated with each frame of input audio features ($\text{x}_{t}$). In addition, we conduct an experiment where the whole-utterance intent embeddings from the final frame ($\text{e}_{T}$) are repeatedly concatenated with all the acoustic feature frames. This serves as an upper bound to the performance of a streaming RNN-T system using an auxiliary A2I model.

\vspace{-6mm}
\subsection{Audio-to-intent model}
\label{sec:A2I model}
\begin{figure}
     \centering
     \begin{subfigure}[b]{0.4\textwidth}
         \centering
         \includegraphics[width=\textwidth]{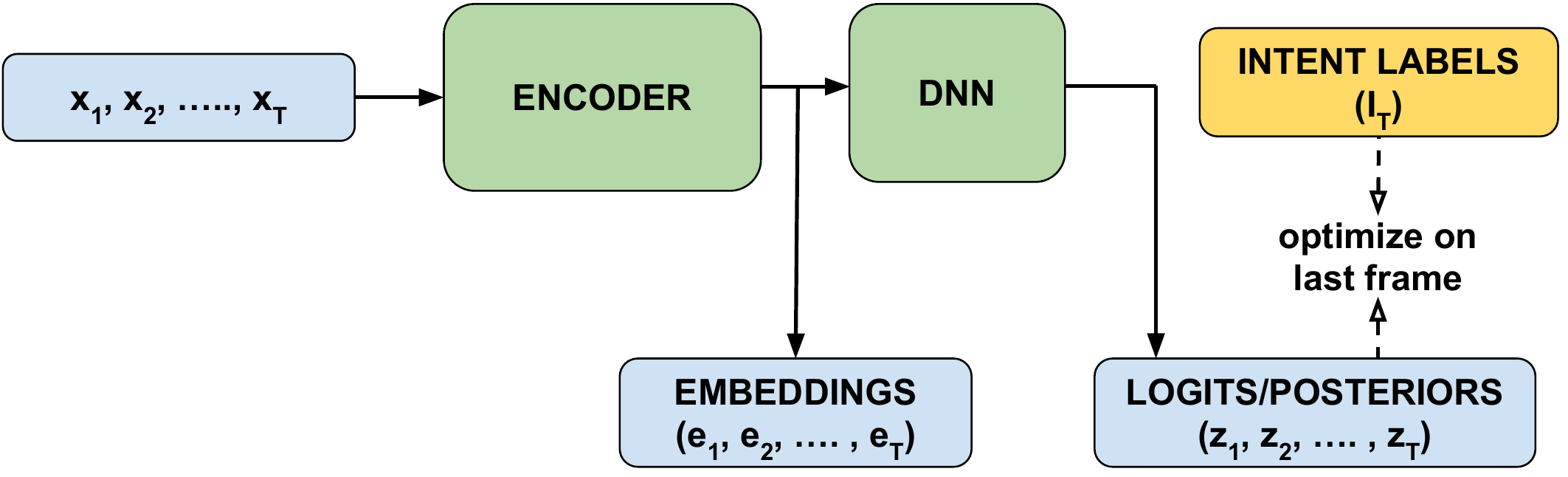}
         \vspace{-4mm}
         \caption{\textit{\footnotesize Optimize intent prediction at only the last frame}}
         \label{fig:A2I-model_final}
     \end{subfigure}
     \hfill
     \begin{subfigure}[b]{0.4\textwidth}
         \centering
         \includegraphics[width=\textwidth]{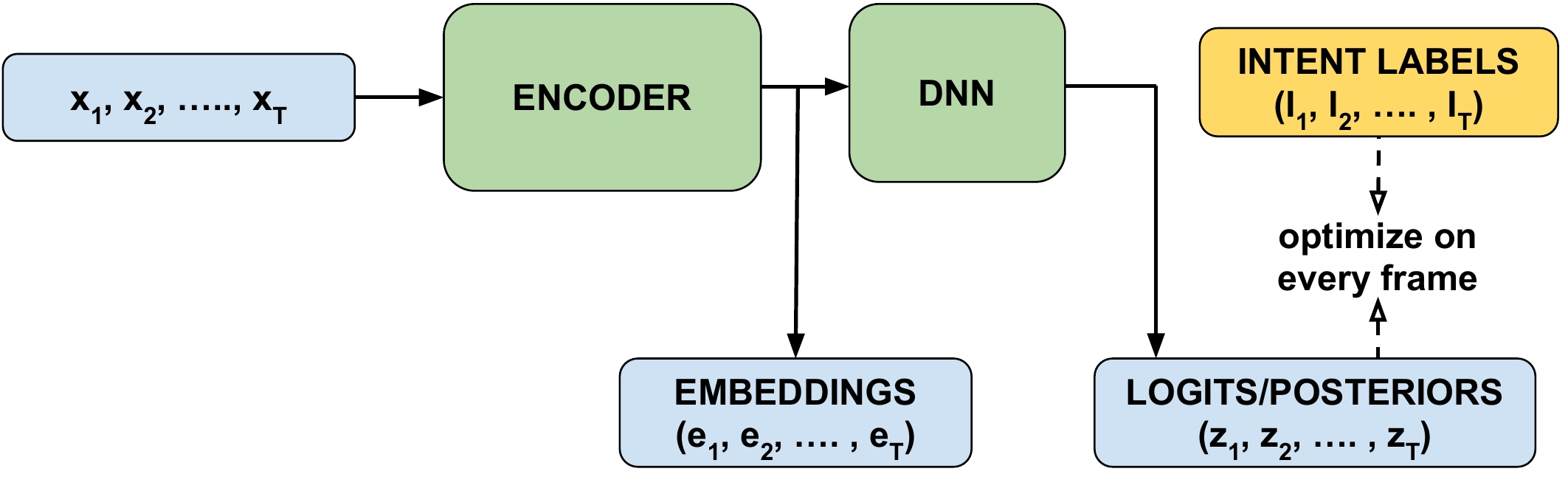}
         \vspace{-4mm}
         \caption{\textit{\footnotesize Optimize intent prediction at every frame; per-frame intent label is the same as the whole-utterance intent label.}}
         \label{fig:A2I-model_every}
     \end{subfigure}
     \vspace{-2mm}
        \caption{\small \textit{Architectures of A2I models}}
        \label{fig:A2I-models}
        \vspace{-4mm}
\end{figure}

Using a separate intent-annotated corpus where each utterance is manually annotated with an intent label, an audio-to-intent (A2I) model can be pre-trained in a supervised manner to classify intent directly from audio input. We explored two A2I models in this work, architectures of which are illustrated in Figure \ref{fig:A2I-models}. In model \ref{fig:A2I-model_final}, frames of audio features ($\text{x}_{t}$) are first encoded into frames of acoustic embeddings ($\text{e}_{t}$) using an LSTM encoder. They are subsequently passed to a dense layer, and intent prediction for the entire utterance is optimized based on posteriors at the last frame. The final-frame intent embeddings $\text{e}_{T}$ can then be concatenated repeatedly with all acoustic feature frames as a non-streaming solution to incorporate intent representation for ASR model building. Model \ref{fig:A2I-model_every} is similar to \ref{fig:A2I-model_final}, except that intent prediction is optimized for each audio frame. The per-frame intent label is the same as the whole-utterance intent label when training model \ref{fig:A2I-model_every}. Per-frame intent embeddings $\text{e}_{t}$ from this model can then be concatenated with each frame of input audio features in a streaming fashion to train our proposed RNN-T ASR model.

\vspace{-1mm}
\section{Data and experimental setup}
\label{sec:setup}

\subsection{Datasets}
\label{sec:datasets}

We use two far-field US English speech datasets in this work to train the A2I and ASR models respectively. The speech data used for training and evaluation are de-identified and based on queries to smart speakers.
\begin{enumerate}
\item{\textit{16k-hour SLU dataset}: Human annotations of NLU intent are available for all utterances. In addition, a development set of 96 hours of data is available. We use this dataset to train the A2I model.}
\item{\textit{50k-hour ASR dataset}: Human transcriptions are available for all utterances. In addition, a test set of approximately 200 hours is available. We use this dataset to train both the baseline and the proposed RNN-T ASR system. Note that 64\% of this dataset contains human intent annotations, which we will use for the partial oracle experiment in Section \ref{sec:partial oracle}}.
\end{enumerate}

\vspace{-2mm}
\subsection{Models}

\subsubsection{RNN-T ASR}
The baseline RNN-T ASR model is trained on the 50k-hour ASR dataset.
The encoder consists of 5 LSTM layers of 1024 hidden units, with a final layer output dimension of 4001. The prediction network has an input embedding layer of 512 units, 2 LSTM layers of 1024 units, and a final output dimension of 4001. The joint network adds the outputs from the encoder and prediction networks as in \cite{graves2012sequence}. These outputs of size 4001 are softmax normalized and correspond to subword tokens of the same vocabulary size. The subword vocabulary was generated using the byte pair encoding algorithm \cite{shibata1999byte}.

\vspace{-2mm}
\subsubsection{A2I models}
\label{sec:model-a2i-models}
A2I models are trained on the 16k-hour SLU dataset with human-annotated intent labels. The model is trained to predict 64 intents, including the 63 most frequent intents and an ``Other'' class that covers all remaining intents. Our A2I models, illustrated in Figure \ref{fig:A2I-models}, use an encoder of 2 LSTM layers with 512 units. The second LSTM layer has a projection layer generating an embedded representation of dimension 64. The projected representation then goes through an extra feed-forward layer of 64 units followed by softmax normalization for the final intent prediction.

\vspace{-2mm}
\subsubsection{RNN-T ASR using intent representations}

In general, we concatenate every original frame of audio features with an intent representation as illustrated in Figure \ref{fig:incorporate-intent-representation} to train our proposed RNN-T ASR model on the 50k-hour ASR dataset.
The intent representations are inferred from the pretrained A2I model, and we freeze weights in the A2I model during the ASR model training.
Note that we do not need NLU annotations for ASR training at this stage as we already have a pretrained A2I model. Thus, our proposed method does not impose any restriction on RNN-T ASR training data.

Since 64\% of the ASR dataset has human-annotated intent labels, we also conduct a partial oracle experiment (Section \ref{sec:partial oracle}) by concatenating either a one-hot representation or an all-zero vector with each frame of input audio features, depending on whether human-annotated intent label for that audio is available.

The audio features used by both A2I and ASR models are 64-dimensional log filter bank energy features computed over a 25ms window with 10ms shift.  Each feature vector is stacked with 2 frames to the left and downsampled to a 30ms frame rate \cite{pundak2016lower}. We use SpecAugment \cite{park2019specaugment}  during RNN-T training to improve model robustness.
All models are trained using the Adam optimizer \cite{kingma2014adam}, with a learning rate schedule including an initial linear warm-up phase, a constant phase, and an exponential decay phase following \cite{chen2018best}. These hyper-parameters are not specifically tuned for this work.

\vspace{-2mm}
\section{Results and discussion}
\label{sec:results-and-discussion}
\subsection{Audio-to-intent accuracy}
\label{sec:A2I-accuracy}

 \vspace{-3mm}
\begin{table}[h!]
    \small
    \caption{\small \textit{{Audio to intent prediction accuracy on 150k samples.}}}
    \vspace{-3mm}
    \centering
    \scalebox{0.9}{
    \renewcommand{\arraystretch}{0.85} 
    \begin{tabular}{l| c |c | c| c }
        \toprule
        	\multirow{2}{*}{A2I model}		 	& \multicolumn{4}{c}{intent prediction accuracy (\%)} \\
	\cline{2-5}
	\multirow{2}{*}{optimized on} &\multicolumn{2}{c|}{train}  &  \multicolumn{2}{c}{dev} \\
         \cline{2-5}
         	 		& per-frame & per-utt& per-frame&per-utt  \\
	
        \midrule
        last frame	& 47.97& 88.92&47.37 &87.70  \\
        every frame	 	& 60.47& 87.16 &59.81 &86.98  \\
        \bottomrule
    \end{tabular}
    }
    \label{tab:A2I-accuracy}
    \vspace{-3mm}
\end{table}
\vspace{-3mm}

\begin{table*}[!h]
    \small
    \caption{\small \textit{Per-intent relative WERR (\%) by using different intent representations (compared to RNN-T ASR baseline).}}
    \vspace{-3mm}
    \centering
    \scalebox{0.9}{
    \renewcommand{\arraystretch}{0.85} 
    \begin{tabular}{l r r r r}

       \midrule
        Intent 	& \multirow{2}{*}{\# utt} 		&  RNN-T + human annotation 			& RNN-T + A2I 				&  RNN-T + A2I  \\
        	 (most frequent from test set)					&		& (one-hot)				& (repeated final embeddings)	& (per-frame posteriors) \\
        \midrule
        None (human annotation not available)				&100.2k		&1.46				&4.88				&3.61 	\\
        PlayMusicIntent			&14.0k		&4.89					&9.34				&9.12	\\
        StopIntent				&8.9k		&23.69					&4.85				&2.52	\\
        ContentOnlyIntent		&5.4k		&14.85					&6.94				&3.45	\\
        YesIntent				&3.6k		&29.31					&11.50				&-4.63	\\
        SetNotificationIntent		&3.5k		&8.31					&4.00				&-3.69	\\ 
        PlayVideoIntent			&3.1k		&8.38					&7.77				&7.93\\
        GetWeatherForecastIntent&2.8k		&5.95					&8.63				&4.17	\\
        PlayStationIntent		&2.7k		&4.67					&7.65				&9.21	\\
        NoIntent				&2.6k		&36.88					&13.66				&3.26	\\
        GetContentIntent		&2.4k		&1.52					&-0.25				&-5.32	\\	
        MusicControlIntent		&2.2k		&24.68					&3.34				&-7.97	\\ 
        \bottomrule
    \end{tabular}
    }
    \label{tab:per-intent-WERR-breakdown}
    \vspace{-2mm}
\end{table*}

\begin{figure}[h!]
     \centering
     \begin{subfigure}[b]{0.45\textwidth}
         \centering
         \includegraphics[width=0.7\textwidth]{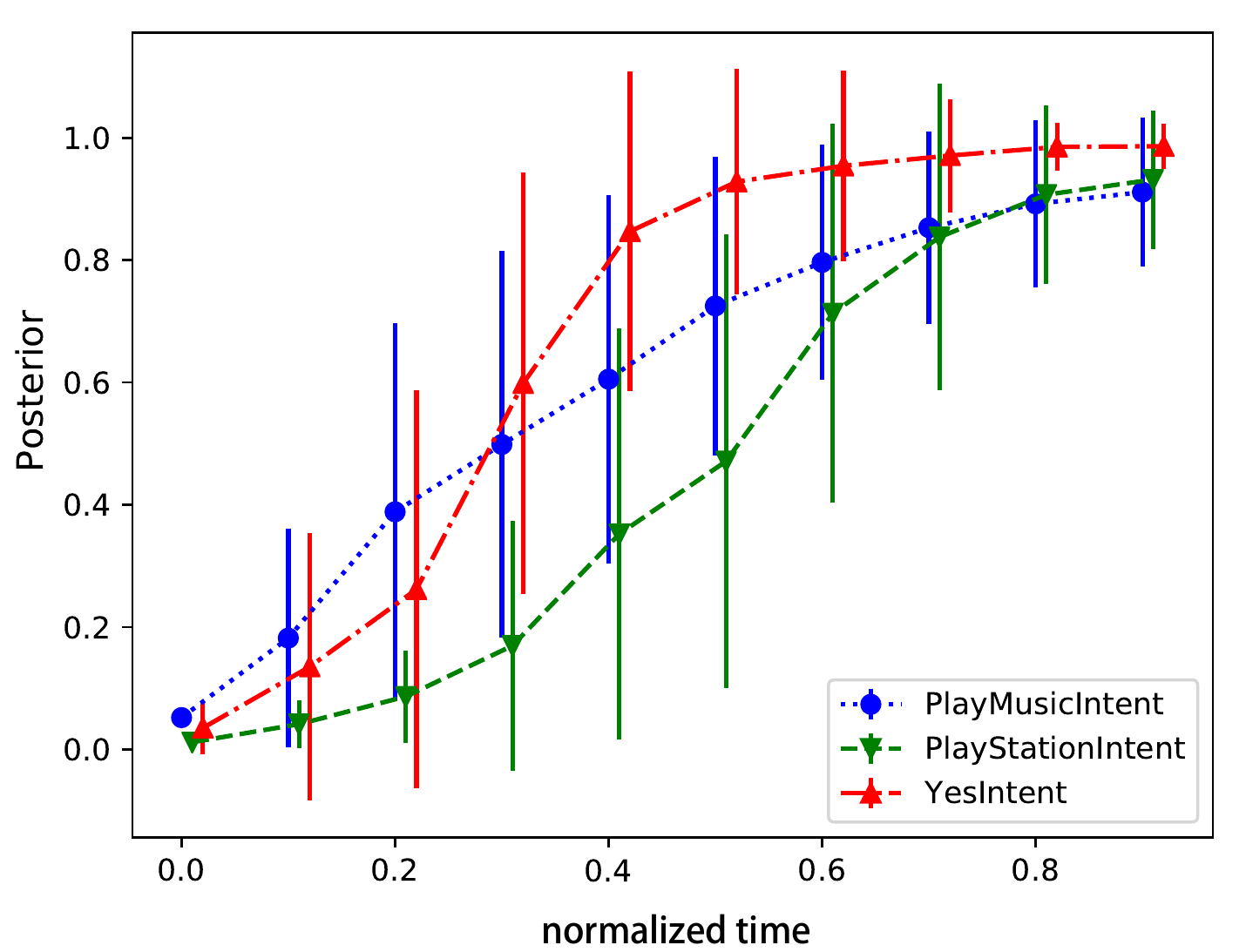}
         \vspace{-2mm}
         \caption{\textit{\footnotesize Mean and standard deviation of posterior over the normalized time for selected intents.}}
         \label{fig:posterior_time}
     \end{subfigure}%
     \hfill
     \begin{subfigure}[b]{0.45\textwidth}
         \centering
         \includegraphics[width=0.8\textwidth]{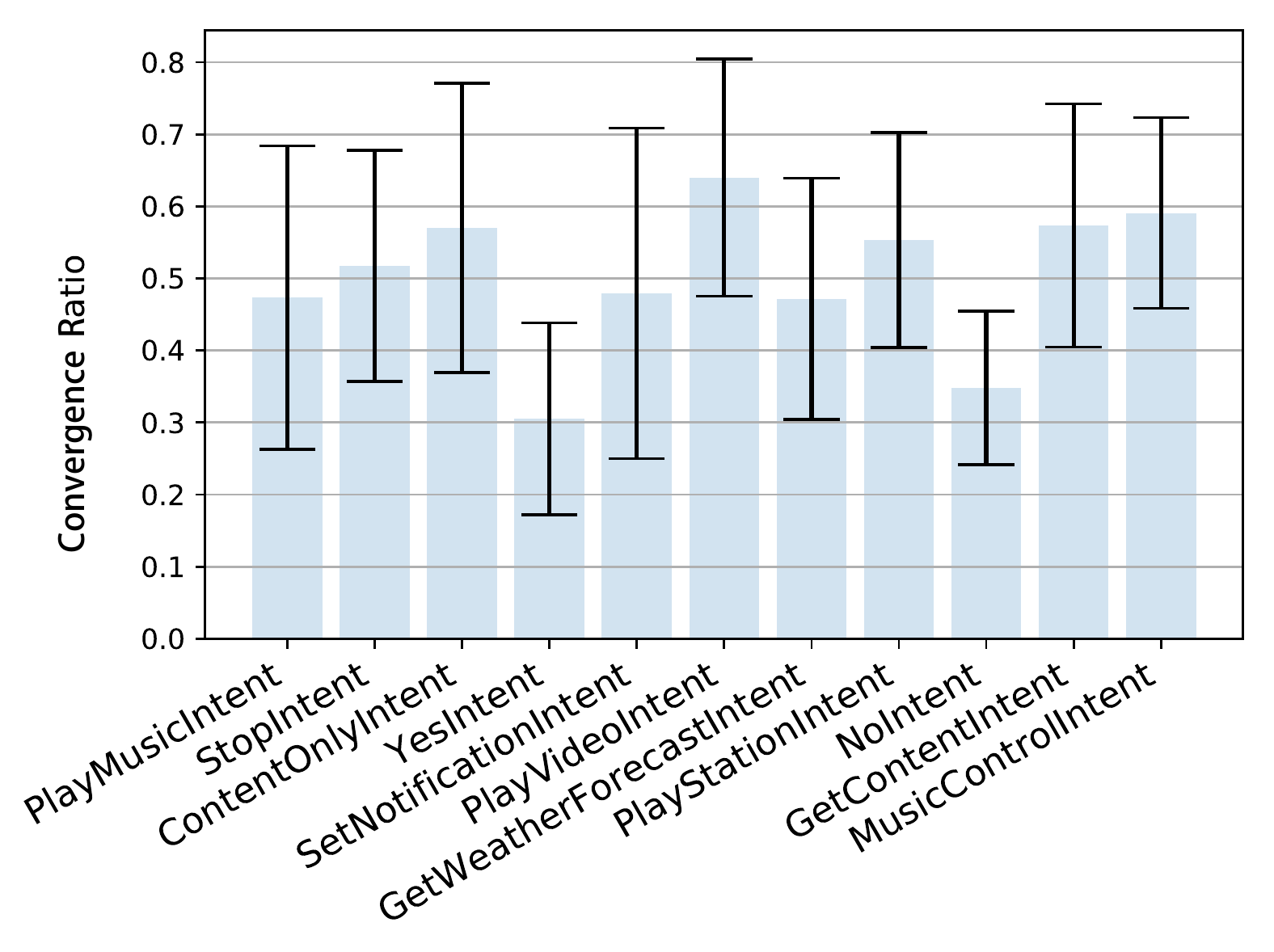}
         \vspace{-3mm}
         \caption{\textit{\footnotesize Mean and standard deviation of convergence ratio for selected intents. The convergence ratio is defined as the normalized time when 80\% of the final-frame intent posterior is achieved.}}
         \label{fig:convergence}
     \end{subfigure}%
     \vspace{-3mm}
        \caption{\small \textit{Per-intent posterior analysis for the A2I model optimized on every frame.}}
        \vspace{-2mm}
        \label{fig:posterior}
\end{figure}

Table \ref{tab:A2I-accuracy} shows the intent prediction accuracy of the two A2I models we have trained. Per-utterance accuracy is calculated based on prediction at the last frame comparing with the whole-utterance NLU intent label from human annotator. Per-frame accuracy is computed based on prediction at every frame and uses the same NLU intent label for all frames.  
Per-frame prediction accuracy is relatively low because intent cannot be predicted reliably in the initial portion of an utterance, but integrating frame-wise intent representation into an ASR system makes the solution streamable.

\begin{table*}[!h]
    \small
    \caption{\small \textit{{Relative WERR (\%) by using different intent representations (compared to RNN-T ASR baseline).}}}
    \vspace{-3mm}
    \centering
    \scalebox{0.9}{
     \renewcommand{\arraystretch}{0.85}
    \begin{tabular}{c|l|l| l |c| c| c }
        \toprule
        No. 	& Exp. 			& A2I model optimized on			&  Intent representation 	& Streamable		&  relative WERR(\%)		& \#params \\
        \midrule
        1	&RNN-T (baseline) 				& - 					& - 					& yes 			& - 		 		& 63.5M\\ 
        2	&RNN-T (larger) 		& - 					& - 					& yes 			& 1.91 		 		& 67.3M\\ 
	3	&RNN-T + A2I				& last frame (Fig. \ref{fig:A2I-model_final})	& repeated final-frame embeddings ($\text{e}_{T}$) 	& no 				&  \textit{5.56} 	& 67.1M\\
	4	&RNN-T + A2I				& last frame (Fig. \ref{fig:A2I-model_final})	& per-frame embeddings ($\text{e}_{t}$)	&yes 		 	& \textit{ 2.78} 		& 67.1M\\
	\midrule
	5	&RNN-T + A2I				& every frame (Fig. \ref{fig:A2I-model_every})	& per-frame embeddings ($\text{e}_{t}$) 	&yes 			& \textit{ 2.89} 	& 67.1M\\
	6	&RNN-T + A2I				& every frame (Fig. \ref{fig:A2I-model_every}) 	& per-frame posteriors ($\text{z}_{t}$)	&yes 			& \textit{ 3.33} 		& 67.1M\\ 
        \bottomrule
    \end{tabular}
    }
    \label{tab:ASR-intent-integration-50khr}
    \vspace{-3mm}
\end{table*}

With the A2I model optimized on every frame, Figure \ref{fig:posterior_time} shows mean and standard deviation of the posterior over the normalized time value (i.e., frame index to total length of utterance). For all utterances with the same predicted final intent, we compute mean and standard deviation of the posteriors at different normalized time stamps.
Figure \ref{fig:convergence} shows the mean and standard deviation of the convergence ratio. This convergence ratio is defined as the ratio of the selected frame index to the total length of utterance, where the selected frame index is the first frame that achieves $80\%$ of the final-frame intent posterior. These statistics are computed over all utterances with correct intent label prediction. The posterior values over time for all utterances are smoothed using a convolution filter with window size 4. These figures show that the confidence of the A2I model increases with more frames and the A2I model can generally predict the true intent from the initial 70\% of the utterance.

\subsection{RNNT-ASR using intent representations}
\subsubsection{Using human-annotated intent labels}
\label{sec:partial oracle}
We conduct a partial oracle experiment by training an RNN-T ASR model with the partially available human-annotated intents in the 50k-hour ASR training dataset. The one-hot intent representation for each frame is concatenated with the original LFBE features as new input to the encoder. We use all-zero vectors when human intent annotation is not available for a given utterance. We believe this acts as a regularizer helping the model to not depend entirely on intent information. However, results reported from this experiment will not reflect the lowest achievable WER since only 64\% of the training dataset contains human intent annotation. Rather, they serve to verify the hypothesis that injecting intent information explicitly into an ASR system can help achieve lower WER. Column 3 of Table \ref{tab:per-intent-WERR-breakdown} reports relative word error rate reduction (WERR) for different intents by using human-annotated intent labels as extra input features. Note that, there is a ``None'' entry since not all utterances have human annotation in the training and test sets, and its relative WERR reduction is relatively small (1.46\% in the table). This is because all-zero vectors are provided as auxiliary inputs for these utterances, which does not provide additional intent-based information for the ASR model to constrain the search space. For test utterances with human-annotated intents, we observe good relative WERRs, especially for those that have limited word choices in utterances (e.g. YesIntent, NoIntent, StopIntent). 

\vspace{-2mm}
\subsubsection{Using Intent Representations from the A2I Model}
For the real use case, we need an A2I model that infers intent representations from audio to help train the ASR model and run the inference, since human-annotated intent labels are not available when decoding input audio after model deployment.

Given the A2I model optimized with last-frame prediction (Figure \ref{fig:A2I-model_final}), we concatenate repeated final frame embeddings ($\text{e}_{T}$) with the original frames of acoustic features  ($\text{x}_{t}$) as new inputs for the encoder of the proposed RNN-T ASR model (row 3 in Table \ref{tab:ASR-intent-integration-50khr}). Since per-frame embeddings ($\text{e}_{t}$) are also available from this A2I model, we concatenate them with the original frames of acoustic features as an extra experiment (row 4 in Table \ref{tab:ASR-intent-integration-50khr}).  Given the A2I model optimized with frame-wise prediction (Figure \ref{fig:A2I-model_every}), we concatenate frame-wise embeddings or posteriors ($\text{e}_{t}$ or $\text{z}_{t}$) with the original frames of acoustic features  ($\text{x}_{t}$) as inputs for the encoder of the RNN-T ASR model (rows 5,6 in Table \ref{tab:ASR-intent-integration-50khr}). Using per-frame embeddings along with the audio feature vectors makes the resulting system fully streamable, while using repeated final embeddings indicates the largest gain achievable by using intent representation as auxiliary input.

Overall, incorporating repeated final-frame embeddings yields a 5.56\% relative WERR while feeding per-frame posteriors in a streaming fashion provides a 3.33\% relative WERR compared to the RNN-T ASR baseline (row 1 in Table \ref{tab:ASR-intent-integration-50khr}). When comparing different intent representations, we see that using per-frame posteriors gives a slightly better relative WERR (3.33\%) compared to using per-frame embeddings (2.89\%). We have also trained a larger RNN-T ASR baseline (row 2 in Table \ref{tab:ASR-intent-integration-50khr}) with its number of parameters comparable to our proposed RNN-T ASR system using an auxiliary A2I model. This larger baseline only yields a 1.91\% relative WERR, which demonstrates that the gain of our proposed method by using the auxiliary A2I model does not come only from the enlarged parameter space.

For the case of using repeated final frame embeddings and per-frame posteriors, we also break down the relative WERR under different intents as shown in Table \ref{tab:per-intent-WERR-breakdown}. Overall, we see gains on various intents.
For the streaming system using per-frame posteriors, which could be applied in a real-time recognition system, we see good relative WERRs in media-playing related intents (9.12\% relative WERR on PlayMusicIntent, 7.93\% relative WERR on PlayVideoIntent and 9.21\% relative WERR on PlayStationIntent). For the unannotated partition (intent ``None"), we see that the RNN-T with A2I model outperforms the system using one-hot representations based on human annotation, which indicates that the A2I model can learn reasonable representations for these data while the one-hot model just treats this case as out-of-domain. This establishes the advantage of using the auxiliary A2I model since it does not need human intent annotation during ASR training and inference. Compared with the non-streamable RNN-T ASR system using repeated final-frame intent embeddings, it seems the streamable RNN-T ASR system using frame-wise posteriors preserves recognition accuracy on intents with longer utterances, such as PlayVideoIntent (3.13k ms on average) and PlayMusicIntent (3.07k ms on average), while its performance degrades on intents with relatively shorter utterances, such as YesIntent (1.37k ms on average) and StopIntent (1.30k ms on average).

Considering the WER gap between the non-streaming system and the streaming system, we perform a diagnostic experiment which feeds per-frame embeddings $\text{e}_{t}$ for a certain percentage of frames in the utterance at the beginning and uses the final-frame embeddings $\text{e}_{T}$ repeatedly afterwards. As shown in Table \ref{tab:semi-oracle}, feeding the final-frame intent embeddings after $70\%$ of the audio only results in slight degradation compared to feeding final-frame intent embeddings from the start of an utterance. This indicates that intent information tends to improve the ASR system more towards the end of the utterance. 

\vspace{-2mm}
\begin{table}[!h]
    \small
    \caption{\small \textit{{Relative WERR (\%) by feeding per-frame embeddings $\text{e}_{t}$ for a certain percentage of frames in the utterance at the beginning and using the final-frame embeddings $\text{e}_{T}$ repeatedly afterwards.}}}
    \vspace{-3mm}
    \centering
    \scalebox{0.9}{
    \renewcommand{\arraystretch}{0.85}
    \begin{tabular}{c c }
        \toprule
       percent (\%) of frames in the utterance  & \multirow{2}{*}{relative WERR (\%)} \\
       after which $\text{e}_{T}$ is fed	&\\
       
        \midrule
       	0 & \textit{5.56}\\
       	50 & \textit{5.44}\\
	70 & \textit{5.00}\\
	100 & \textit{2.78}\\
        \bottomrule
    \end{tabular}
    }
    \label{tab:semi-oracle}
    \vspace{-3mm}
\end{table}

\section{Conclusions}
\label{sec:conclusion}
We have demonstrated the benefit of incorporating intent representations for improving end-to-end ASR system with RNN-T. We showed that an audio-to-intent (A2I) auxiliary model is capable of predicting intent representations that substantially help lower ASR WER. Over all intent classes, incorporating repeated whole-utterance intent representation in a non-streaming fashion gives a 5.56\% relative WERR, while feeding per-frame intent posteriors in a streaming fashion brings a 3.33\% relative WERR. In the future, we hope to investigate the effectiveness of our approach as a function of NLU and ASR data size, as well as incorporating other types of auxiliary information. 

\newpage

\bibliographystyle{IEEEtran}

\bibliography{refs}

\end{document}